\documentclass[a4paper,onecolumn,conference]{IEEEtran}

\usepackage{ifpdf}
\usepackage{cite}
\usepackage[pdftex]{graphicx}
\usepackage{array}
\usepackage{mdwmath}
\usepackage{mdwtab}
\usepackage{amssymb,latexsym}
\usepackage{stfloats}
\usepackage{amsmath}
\usepackage{subfig}
\usepackage{algcompatible}
\usepackage{algorithm}
\usepackage[font={footnotesize}]{caption}
\usepackage{balance}
\usepackage[utf8]{inputenc}
\usepackage[T1]{fontenc}
\usepackage[compatible]{algpseudocode}
\usepackage{hyperref}
\usepackage{footnote}

\renewcommand{\figurename}{Figure}

\DeclareRobustCommand*{\IEEEauthorrefmark}[1]{%
\raisebox{0pt}[0pt][0pt]{\textsuperscript{\footnotesize\ensuremath{#1}}}}

\renewcommand\IEEEkeywordsname{Keywords}

\begin{document}

\title{Epicardial Adipose Tissue Segmentation From CT Images With A Semi-3D Neural Network} 
\author{\IEEEauthorblockN{
Marin Benčević\IEEEauthorrefmark{1},
Marija Habijan\IEEEauthorrefmark{1},
Irena Galić\IEEEauthorrefmark{1}}
\IEEEauthorblockA{\IEEEauthorrefmark{1}
Faculty of Electrical Engineering, Computer Science and Information Technology Osijek\\
Kneza Trpimira 2b, Osijek, 31000, Croatia}
{\it marin.bencevic@ferit.hr}}

\maketitle

\begin{abstract}
Epicardial adipose tissue is a type of adipose tissue located between the heart wall and a protective layer around the heart called the pericardium. The volume and thickness of epicardial adipose tissue are linked to various cardiovascular diseases. It is shown to be an independent cardiovascular disease risk factor. Fully automatic and reliable measurements of epicardial adipose tissue from CT scans could provide better disease risk assessment and enable the processing of large CT image data sets for a systemic epicardial adipose tissue study. This paper proposes a method for fully automatic semantic segmentation of epicardial adipose tissue from CT images using a deep neural network. The proposed network uses a U-Net-based architecture with slice depth information embedded in the input image to segment a pericardium region of interest, which is used to obtain an epicardial adipose tissue segmentation. Image augmentation is used to increase model robustness. Cross-validation of the proposed method yields a Dice score of 0.86 on the CT scans of 20 patients.
\end{abstract}


\let\thefootnote\relax\footnotetext{This is the accepted version of a paper published at
  2021 International Symposium ELMAR 978-1-6654-4437-8/21/\$31.00 ©2021 IEEE;\\
  DOI: 10.1109/ELMAR52657.2021.9550936; \url{https://ieeexplore.ieee.org/abstract/document/9550936}.}

\IEEEpeerreviewmaketitle

\section{Introduction}
\label{intro}

Epicardial adipose tissue (EAT) is a type of adipose tissue located within the pericardium, a protective layer around the heart. EAT lies between the myocardium (the heart muscle) and the fibrous outer layer of the pericardium. Due to EAT's proximity to the myocardium, it is believed to be a metabolically active organ that has a direct impact on various cardiovascular diseases (CVDs).

EAT has been shown to play a direct role in coronary atherosclerosis and cardiomyopathy \cite{Sacks2007, Marwan2013}. EAT thickness has been shown to correlate with metabolic syndrome \cite{Chenn2009} and coronary artery disease independently of obesity \cite{Iacobellis2011}. It relates in general with the progression of coronary artery calcification \cite{Mahabadi2014, Gorter2008}. Additionally, the volume and density of EAT have been linked to major adverse cardiac events in asymptomatic subjects \cite{Goeller2018}. EAT volume is also a good marker for cardiovascular diseases \cite{Raggi2013, Mahabadi2013}. Additionally, EAT plays a role in insulin resistance, is an accurate therapeutic target, and impacts heart morphology and adiposity \cite{Iacobellis2009-2}.

EAT's active role in CVDs makes it a valuable diagnostic tool. Therefore, measuring EAT volume and thickness is an important medical task. Currently, EAT is most often quantified by measuring EAT thickness using echocardiography. Estimating volume from thickness at a single point can introduce inaccuracies and inter-observer variability. Using 3D medical imaging technologies such as computerized tomography (CT) results in more accurate measurements, however, the availability of CT machines as well as the procedure's cost and duration make it impractical for common clinical use. One downside of using CT to measure EAT is the time required to manually measure EAT volume. Manually quantifying EAT from CT images can take up to an hour per patient \cite{Militello2019} and is prone to inter-observer variability of around 10\% \cite{Marwan2013} of EAT volume. Fully or semi-automatic methods of EAT segmentation and quantification from CT images could reduce time requirements and the cost of EAT quantification.

Segmenting EAT is a challenging image processing task. Its uneven distribution around the heart, peculiar shape, and its similarity to other adipose tissues nearby complicate the process. Segmenting EAT relies on delineating the pericardium, which is less than 2 mm thin and can often be hard to delineate on CT images due to partial volume effects.

\subsection{Related Work}

There are several semi- and fully automatic methods proposed in the existing literature. One of the first methods of EAT segmentation was proposed by Coppini et al. \cite{Coppini2010}. They use a semi-automatic approach where an expert places control points on the pericardium which are used to define a volume of interest (VoI). The VoI is then thresholded to the adipose tissue range, and the segmentation is refined using geodesic active contours. An improved semi-automatic segmentation method is presented by Militello et al \cite{Militello2019}. Their method requires expert input to define the VoI on a few slices of the scan, and the rest of the VoI is then interpolated, saving time while still offering manual-level segmentation accuracy. 

Ding et al. \cite{Ding2014} propose a method similar to Coppini et al. \cite{Coppini2010} but instead of relying on manual initialization, they initialize the pericardium contour using an atlas-based method. They then use geodesic active contours to refine the contour and obtain EAT volume by thresholding the region inside the pericardium contour. Shahzad et al. \cite{Shahzad2013} propose another fully automatic method. They use multi-atlas pericardium segmentation. The volume inside the pericardium is then thresholded, followed by a connected component analysis step to remove noise from the final EAT volume. Rodrigues et al. \cite{Rodrigues2016} propose a fully automatic machine-learning-based method. Their method consists of extracting hand-selected salient features from the CT slices, which are then segmented by a learned random forest classifier.

Commandeur et al. \cite{Commandeur2018} proposed one of the first EAT segmentation methods to use deep learning. They use a multi-task approach consisting of two convolutional neural networks (CNNs). The networks first classify whether a slice contains the heart, and then segment, in parallel, (1) the union of cardiac adipose tissue and (2) EAT and other cardiac adipose tissues separately. They also employ statistical shape modeling to refine the segmentation results. More recent works like Li et al. \cite{Li2019} and He et al. \cite{he2020}  use deep learning based on variations of the U-Net architecture.

This paper presents a fully automatic EAT segmentation method to aid in giving accurate measurements of EAT from CT scans. We use a U-Net-based \cite{ronneberger2015unet} architecture to segmentent a pericardium region. This region is then used to threshold the input image to obtain a final EAT segmentation. While our method works on a per-slice basis, we incorporate slice depth information as a separate channel in each input image, thus utilizing spatial information while maintaining a 2D-based architecture.

The rest of this paper is structured as follows. Section \ref{data} describes the used dataset and preprocessing steps. Section \ref{method} provides an overview of the used method for EAT segmentation. The results of our method are presented in Section \ref{experiment}. Finally, we give a conclusion in Section \ref{conclusion}.

\section{Dataset description}\label{data}

This paper uses a dataset of CT scans of 20 patients from Rio de Janeiro obtained and released publicly by Rodrigues et al. \cite{Rodrigues2016}. The original ground truth was obtained via manual segmentation of 878 total slices by a physician and a computer scientist. The slices were also registered, scaled, and cropped to a similar anatomical region, as well as thresholded to the adipose tissue range of $[-200, -30]$ HU. Of the 20 patients, 10 are male and 10 are female. The mean age of the patients is 55.4. Each scan has an average of 42.95 slices with a slice thickness of 3 mm. The CT scans were obtained with two different scanners, 9 patients were scanned by a Phillips scanner, while the other 11 were scanned by a Siemens scanner.

\begin{figure}[h]
    \centering
    \includegraphics[width=0.6\columnwidth]{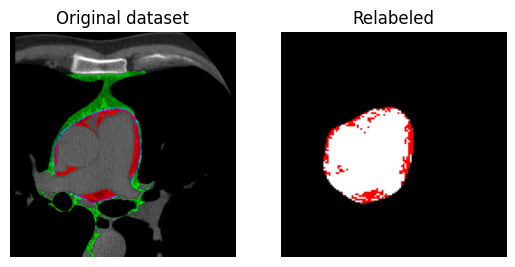}
    \caption{An example of the original image label and our relabeled image. The EAT label is shown in red, while the pericardium label is shown in white.}
    \label{fig:dataset-relabel}
\end{figure}

The original labels contain three classes: pericardium, EAT and paracardial adipose tissues. Using these labels as a reference, we manually labeled a closed pericardium region on each slice. During labeling, we follow the pericardium line where available and mark the pericardium region as the border separating the two adipose tissues when a label is not visible. We then create a separate set of EAT labels by multiplying the input images, thresholded to the range of adipose tissue, with the pericardium label. I.e. we define EAT as all adipose tissue masked by the pericardium label. A comparison of the original label and our re-labeled images is shown in Fig. \ref{fig:dataset-relabel}.

\section{Methodology}\label{method}

\begin{figure}[b]
\center
\includegraphics[width=0.6\columnwidth]{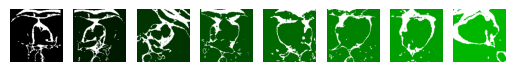}
\caption{A sample of the inputs to the neural network from a single patient, sorted by slice depth. The first channel (CT adipose tissue) is shown in full white, while the second channel (the slice depth) is shown in green.}
\label{fig:input-images}
\end{figure}

We achieve segmentation of the pericardium using a deep neural network architecture based on U-Net \cite{ronneberger2015unet}. The input to the model is a 2-channel $128 \times 128$ image. The first channel is a single slice of a patient's CT scan. The second channel represents the slice depth of that slice. The pericardium region varies highly by slice depth. Therefore, we utilize depth information by first normalizing the slice depth of each input slice to a value between 0 and 1. We then create a $128 \times 128$ image where each pixel has the value of the slice depth. This image is used as the second channel of the neural network input. This allows the network to utilize additional depth information without the need to change the underlying architecture. Example input images are shown in Fig. \ref{fig:input-images}.

We use a loss function which is a modified version of the Dice coefficient, shown in (\ref{eq:loss}).

  \begin{equation}
    \textit{DSC}_{loss} = 1 - \frac {2|X\cap Y| + \lambda}{|X|+|Y|+\lambda}
    \label{eq:loss}
  \end{equation}
  
Where $X$ and $Y$ are the input and predicted images, respectively, and $\lambda$ is a smoothing parameter set to 1 in our experiments.

\subsection{Data preprocessing}

We apply several data preprocessing steps on the original dataset to achieve better segmentation results. First, the images are all normalized and zero-centered by subtracting the global mean intensity of all pixels of the training dataset (0.1). The images are then scaled down from $512 \times 512$ to $128 \times 128$ pixels. We also removed a total of 112 slices from the original dataset that did not include EAT labels, either because of labeling errors in the original dataset or because the slices were outside the heart region.

\begin{figure}[h]
\center
\includegraphics[width=0.6\columnwidth]{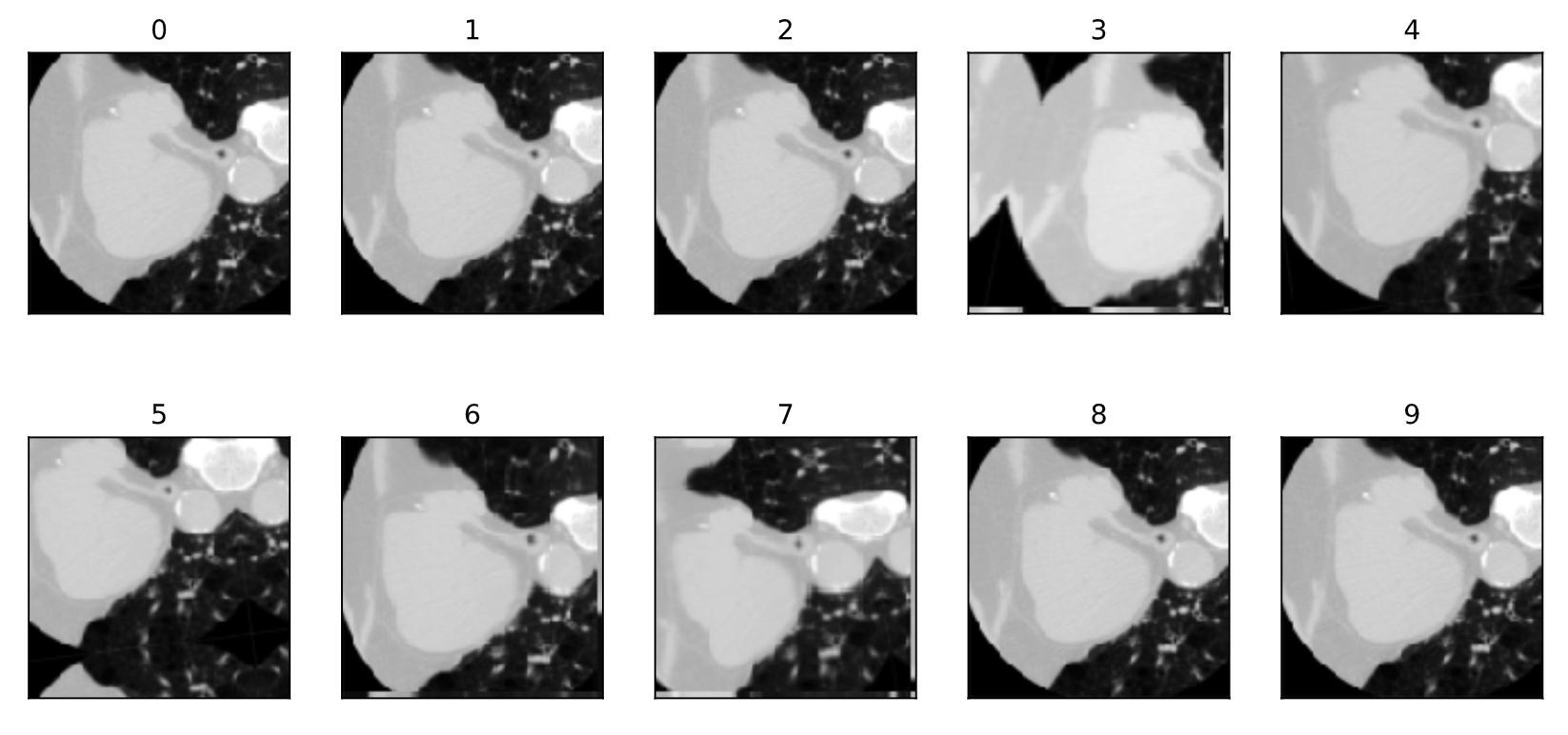}
\caption{Different random augmentation examples of the same input image.}
\label{fig:augment}
\end{figure}

Furthermore, we utilize heavy data augmentation to increase model generalizability. In real-world distributions, anatomical structures can differ drastically from one person to the next. To simulate these differences, we add a random chance of augmenting each input image during the training phase. Each input image has a 50\% chance for a horizontal flip, 30\% chance of a random combination of affine image transformations, including (1) a translation of max. 6.25\% of the image's width, (2) a scaling of max. 10\% of the image's scale, and (3) a rotation of a max. 45 degrees. Additionally, each input image has a 20\% chance of a non-linear mesh deform. Examples of augmented images are shown in Fig. \ref{fig:augment}.

\subsection{Model training}

The neural network is implemented and trained using PyTorch 1.7.1 on an NVIDIA GeForce RTX 3080 GPU. The networks were trained for 200 epochs, with a checkpoint mechanism after each epoch. We select the model with the best validation loss during training. In our experiments, the model converged around epoch 100 after 5 minutes. The training was done using the Adam optimizer with a learning rate of 0.001. The used batch size was 8. We used a manual random seed value of 42 for all experiments. We use 2-fold per-patient cross-validation for all experiments. Each model was trained on the slices of 10 patients and validated on the remaining 10 patients.

\section{Experiments and results}\label{experiment}

The results in this section are evaluated on the two cross-validation folds and averaged across folds. Segmentation quality is evaluated using the Dice coefficient (DSC) and Jaccard index. We analyze the model's quantification quality with a Bland-Altman analysis \cite{MARTINBLAND1986307} as well as with the Pearson correlation coefficient. To calculate the Pearson correlation and to perform the Bland-Altman analysis, we threshold all segmented pericardium regions to obtain the EAT segmentation results and calculate the number of pixels labeled as EAT on each image. We also calculate the number of EAT pixels on the ground truth images. The pixel counts are used as a proxy for volume measurement, as EAT volume of a patient is proportional to the number of segmented EAT pixels in the patient's CT scan.

\subsection{Results}

\begin{figure}[h]
    \centering
    \includegraphics[width=0.6\columnwidth]{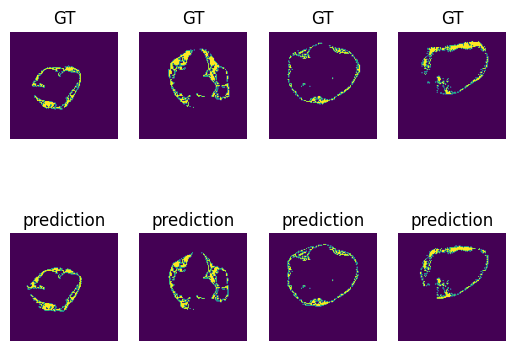}
    \caption{Examples of EAT predictions compared to the ground truth images.}
    \label{fig:examples}
\end{figure}

The DSC and correlation results of the three models are presented in Table \ref{tab:results}. Examples of predictions from our model are presented in Fig. \ref{fig:examples}.

\begin{table}[h]
\renewcommand{\arraystretch}{1.4}
\caption{Mean results of the cross validation. The Corr. value is the Pearson correllation between the total number of EAT pixels for each slice of the validation dataset ($p<0.0001$).}
\centering
\begin{tabular}{cccccc} 
 \hline
 Target & DSC & Jaccard & Prec. & Rec. & Corr. \\
 \hline
 Pericardium & 0.9264 & 0.8819 & 0.9319 & 0.9345 & - \\ 
 EAT & 0.8646 & 0.7807 & 0.8787 & 0.8690 & 0.8864 \\
 \hline
\end{tabular}
\label{tab:results}
\end{table}

\begin{figure}[h]
    \centering
    \includegraphics[width=0.6\columnwidth]{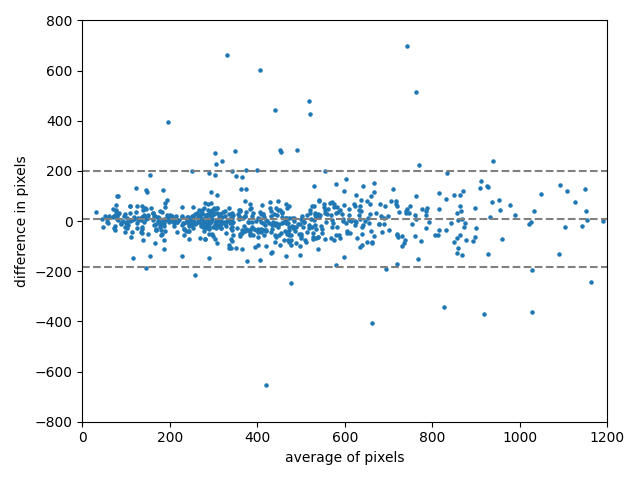}
    \caption{The Bland-Altman analysis of the number of pixels predicted as EAT on each slice of the test dataset for each fold. The dashed lines indicate a 95\% confidence interval.}
    \label{fig:corr}
\end{figure}

The Bland-Altman analysis of our method is presented in Fig. \ref{fig:corr}. The analysis shows a high level of agreement between our method and ground-truth annotations. The plot also shows a small positive bias from our method. Additionally, the plot does not show a strong proportional bias. However, while most measurements fall within the 95\% confidence interval, there are outliers. The most likely reason for these errors is the noisy nature of the ground truth labels especially towards the edges of the heart region.

Additionally, we compare our results with using a U-Net to directly segment the pericardium, as well as with other state-of-the-art approaches This comparison is presented in Table \ref{tab:comparison}.

\begin{table}[h]
\renewcommand{\arraystretch}{1.4}
\caption{A comparison of our approach with other deep-learning-based approaches for EAT segmentation.}
\centering
\begin{tabular}{cccccc} 
 \hline
 Method & DSC & Jaccard & Prec. & Rec. & Params \\
 \hline
 U-Net for EAT & 0.75 & 0.58 & 0.72 & 0.69 & 5.8 M \\ 
 Zhang et al. \cite{Zhang2020} & \textbf{0.91} & \textbf{0.84} & - & - & 11.6 M \\
 He et al. \cite{he2020} & 0.85 & - & 0.86 & 0.89 & 6.4 M \\
 Our method & 0.86 & 0.78 & 0.89 & 0.87 & 5.8 M \\
 \hline
\end{tabular}
\label{tab:comparison}
\end{table}

\section{Conclusion}\label{conclusion}

Our method achieves a mean DSC of 0.8574 and a correlation coefficient of 0.8864. While the results are worse than the current state of the art, our method uses a more direct neural network with fewer preprocessing steps and fewer parameters than deep-learning-based solutions for this task (\cite{Commandeur2018, Li2019, he2020}), leading to lower inference times. We show that it is possible to obtain good results by segmenting the pericardium region instead of EAT directly. The pericardium region has a smooth closed contour, a much easier task to learn when compared to the unevenly and discontinuously distributed EAT. Additionally, we investigate a way to utilize depth information in an encoder-decoder type neural network without the need to train on patched 3D volumes. We embed the slice depth as an additional channel in the image and show that this additional channel improves segmentation results.

The Bland-Altman plot shows the potential of correcting the quantification based on the estimated bias of the model, thus significantly improving quantification performance. This approach is sometimes called "adjusted classify and count", as opposed to a method of simply counting the instances, referred to as "classify and count" \cite{forman05}. With access to more and better quality training data and using the "adjusted classify and count" method, our method could provide significantly improved EAT quantification results.

\clearpage

\section*{Acknowledgment}

This work has been supported in part by the Croatian Science Foundation under the project UIP-2017-05-4968.

\bibliographystyle{ieeetr}
\bibliography{reference}

\end{document}